\documentclass[Instruments,article,submit,moreauthors,pdftex,10pt,a4paper]{mdpi}

\firstpage{1} 
\articlenumber{x}
\doinum{10.3390/------}
\pubvolume{xx}
\pubyear{2016}
\copyrightyear{2016}
\history{Received: date; Accepted: date; Published: date}

\pdfoutput=1

\Title{A Novel Cosmic Ray Tagger System for Liquid Argon TPC Neutrino Detectors}

\Author{M.~Auger$\,^{a}$,
M.~Del Tutto$\,^{b}$,
A.~Ereditato$\,^{a}$,
B.~Fleming$\,^{c}$,
D.~Goeldi$\,^{a}$,
E.~Gramellini$\,^{c}$,
R.~Guenette$\,^{b}$,
W.~Ketchum$\,^{d}$,
I.~Kreslo$\,^{a}$,
A.~Laube$\,^{b}$,
D.~Lorca\thanks{Corresponding author}~$\,^{a}$,
M.~Luethi$\,^{a}$,
C.~Rudolf~von~Rohr$\,^{a}$,
J.~R.~Sinclair$\,^{a}$,
S.~R.~Soleti$\,^{b}$,
M.~Weber$\,^{a}$}  

\address{%

$^{a}$~Laboratory for High Energy Physics, Albert Einstein Center for Fundamental Physics, University of Bern, CH-3012, Switzerland\\

$^{b}$~University of Oxford, Oxford, OX1 3RH, United Kingdom\\

$^{c}$~Yale University, New Haven, CT, 06520, USA\\

$^{d}$~Fermi National Accelerator Laboratory, Batavia, Illinois 60510, USA\\}

\corres{E-mail: david.lorca@lhep.unibe.ch}

\abstract{The Fermilab Short Baseline Neutrino (SBN) program aims to observe and reconstruct thousands of neutrino-argon interactions with its three detectors (SBND, MicroBooNE and ICARUS-T600), using their hundred of tonnes Liquid Argon Time Projection Chambers to perform a rich physics analysis program, in particular focused in the search for sterile neutrinos. Given the relatively shallow depth of the detectors, the continuos flux of cosmic ray particles which crossing their volumes introduces a constant background which can be falsely identified as part of the event of interest. Here we present the Cosmic Ray Tagger (CRT) system, a novel technique to tag and identify these crossing particles using scintillation modules which measure their time and coordinates relative to events internal to the neutrino detector, mitigating therefore their effect in the event tracking reconstruction.}


\begin{document}



\section{Introduction} \label{sec:Introduction}

The Fermilab Short Baseline Neutrino (SBN) program was proposed by three international collaborations to accomplish a wide spectrum of physics research. Once the program is completed, it will perform sensitive searches for $\nu_{e}$ appearance and $\nu_{\mu}$ disappearance, it will also study the neutrino-argon cross section with millions of interactions using the Booster Neutrino Beam~(BNB) at Fermilab~\cite{Antonello:2015lea}.  These studies are expected to solve the currently unexplained low energy event excess observed by LSND~\cite{Aguilar:2001ty} and later confirmed by MiniBooNE~\cite{Aguilar-Arevalo:2013pmq}, which could be interpreted by theories that extend the Standard Model of Particles and Interactions with additional "sterile" neutrinos or the presence of a non-observed-yet electromagnetic background.

The program is composed of the currently running MicroBooNE detector, a 170~t Liquid Argon~(LAr) Time Projection Chamber~(TPC) with charge-collection readout~\cite{Terao:2015sqa}, located 470~m away from the BNB primary target. Besides this, the 112~t Short Baseline Near Detector~(SBND)~\cite{McConkey:2015pba} and 600~t ICARUS T-600 detector~\cite{Montanari:2015zqy}, both LArTPCs, are respectively located at 110~m and 600~m away from the BNB target. The construction of the SBN detectors is expected to be completed in 2017. The MicroBooNE collaboration finished the commissioning of the detector in the Summer of 2015 and has already collected neutrino interactions corresponding to about 4.0$\times$10$^{20}$ protons on target, more than half of the total expected for three years running.

Due to the shallow depth of the detectors' location ($\sim$6 meters underground), these are continuously exposed to a flux of background cosmic ray particles. At ground level, the most abundant of these cosmic ray particles are muons, produced in the decay of pions and kaons~\cite{Surdo:2015tva}.

In a typical neutrino-argon interaction in the LArTPCs, charged particles release their energy in the medium as described by the Bethe-Bloch formula. Due to the electric field present in the TPC, ionization electrons drift toward the anode where they are collected and the energy and momentum of the secondary particles is reconstructed. At the same time, primary scintillation light coming from the argon atoms is readout by a system of photomultiplier tubes immersed in the LAr volume, providing timing information and an external trigger for the data acquisition (DAQ).

There exists a non-zero probability that a high energy cosmic ray muon crossing the reconstructed volume is misidentified as part of the neutrino interaction. Each muon track is surrounded by tracks of electrons and positrons originating from bremsstrahlung of delta-electrons. It has been estimated that around 8~muons cross the 2.33~m~$\times$~2.56~m~$\times$~10.37~m active volume in a 2.2~ms DAQ window of MicroBooNE.

The Cosmic Ray Tagger (CRT) system presented here detects cosmic ray muons and measures their crossing time and coordinates relative to events internal to the TPC. It is a tool to mitigate the cosmic ray background in the MicroBooNE and SBND detectors and to improve the statistical significance of the physics measurements. The CRT systems will enable the MicroBooNE and SBND experiments to efficiently remove cosmogenic related activity from beam neutrino datasets, as well as allowing precision detector response characterization and calibration utilizing tagged cosmic muons.  

We report here on the design and construction features of the CRT basic modules, as well as presenting performance results.

\section{Cosmic Ray Tagger System } \label{sec:CRTdesign}

The CRT system is composed of individual scintillator modules readout by Silicon PhotoMultipliers (see Section~\ref{subsec:ScintMod}) with a maximum size of 4.1~m $\times$ 1.8~m $\times$ 2~cm and a maximum weight of 174~kg. Smaller-sized modules, 25\%~-~80\% of the maximum size, will also be utilized to match to the specific geometry of the experiments at Fermilab. The module detection efficiency for muons has been measured to be higher than 95\% with a coordinate resolution better than 1.8~cm (see Section~\ref{sec:Results}).

The modules are produced at the Laboratory for High Energy Physics (LHEP) in Bern and are shipped to Fermilab after a quality inspection. After assembly, modules and readout electronics (see Section~\ref{subsec:FEB}) are tested for proper operation. These tests verify the full readout chain, characterize the efficiency of cosmic muon detection in each module, and measure the data collection processing in each channel.

\subsection{Scintillator Module} \label{subsec:ScintMod}

\begin{figure}[h!]
\centering	
\includegraphics[trim = 10mm 10mm 10mm 40mm, clip, width=0.78\textwidth, angle=0]{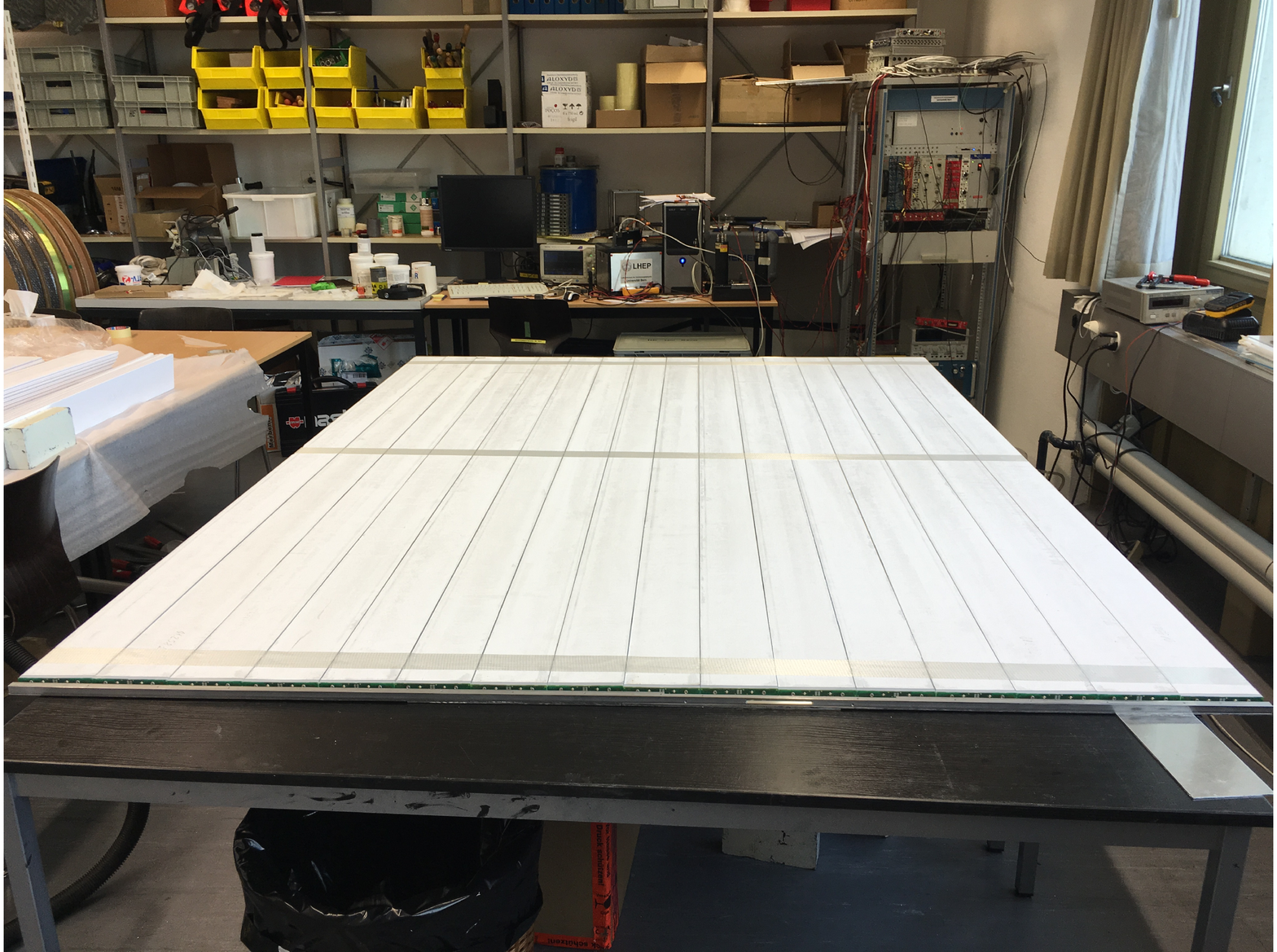}
\caption{Scintillator module containing 16 scintillating strips. The external case is still to be mounted.}
\label{Plane1}
\end{figure}

Each scintillator module consists of a row of 16 mechanically joined 10.8~cm wide scintillator strips (Figure~\ref{Plane1}) placed side by side in a protective aluminum casing. The thickness of the case walls is 2 mm; this provides the necessary mechanical stability. Strips are fixed within the module by a 0.1~mm thick double-sided adhesive layer.  The gap between two adjacent strips in a module is kept below 0.2~mm. The length of the modules varies from 2.3~m to 4.1~m, for three different widths: 0.97~m, 1.75~m and 1.81~m.

\subsection{Scintillator Strip}

\begin{figure}[h!]
\centering	
\includegraphics[trim = 0mm 20mm 0mm 20mm, clip, width=0.99\textwidth]{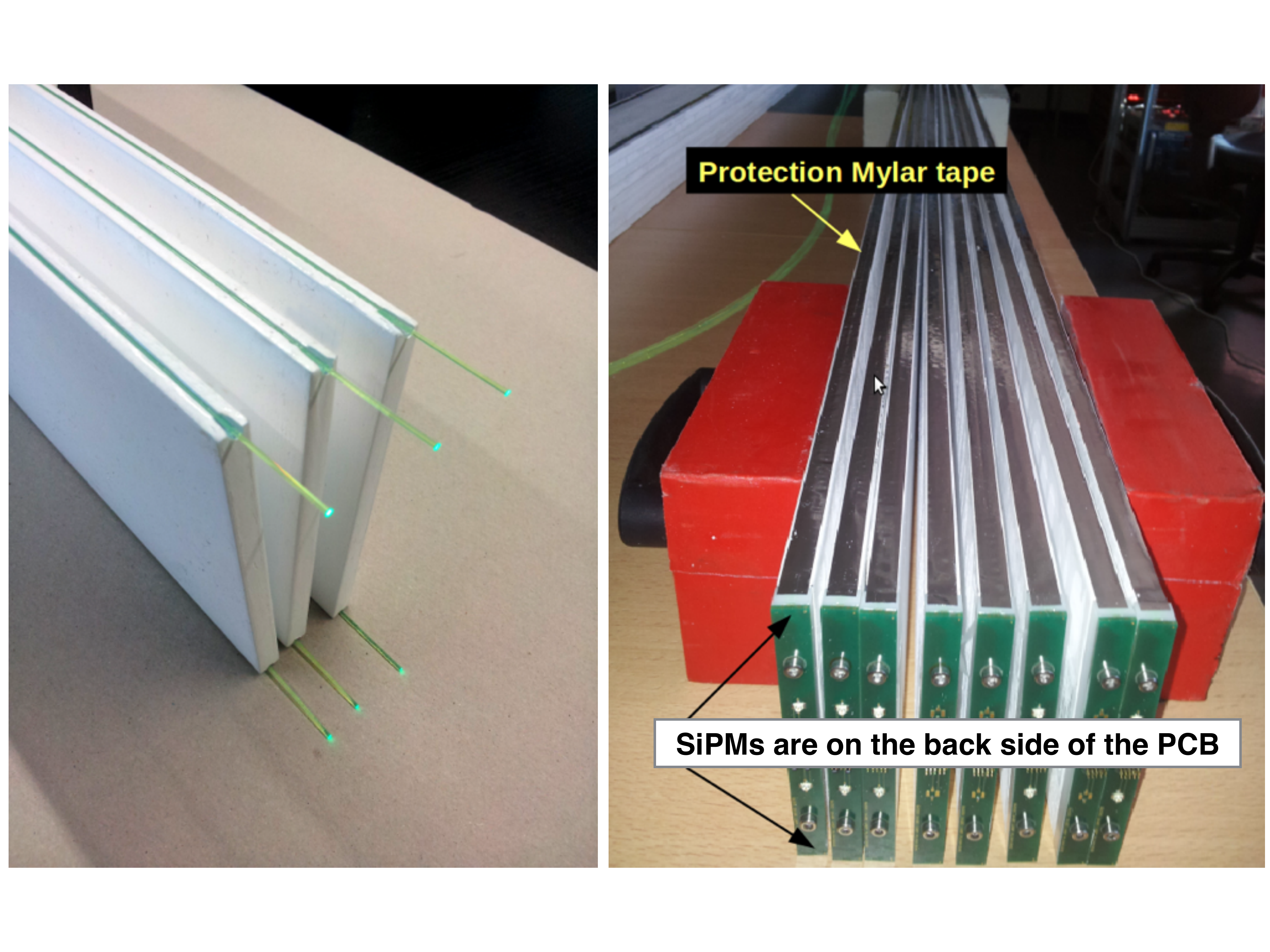}
\caption{Scintillator strips with wavelength shifter fibers glued into grooves along the strip long edges. The strip is covered by a high-reflectivity white layer. The scintillation light is collected by Kuraray Y-11 wavelength shifting fibers and transmitted to the strip end where it is detected by Hamamatsu S12825-050P SiPMs.}
\label{Strip}
\end{figure}

Scintillating strips are extruded from a USMS-03\footnote{USMS-03 by UNIPLAST Inc, Vladimir, Russian Federation.} polystyrene-based mixture containing 1.5\% of Diphenylbenzene (PTP) and 0.04\% of Bis(5-phenyl-2-oxazolyl)benzene (POPOP). The surface of the strips is subjected to structural modification resulting in the formation of a highly-reflective white layer (Figure~\ref{Strip}-left). 

The scintillator has an emission maximum at 430~nm and a measured bulk attenuation length of longer than 7.5~cm. In order to provide efficient and uniform collection of the scintillation light two wavelength shifting (WLS) fibers (Kuraray Y11(200)M, 1 mm diameter) are glued into two grooves in the long edges of each strip with a polysiloxane compound. The high elasticity of the compound allows a displacement of the fiber along the strip without the risk of damaging its cladding.  This, in turn, adds to the robustness of the assembled module. The modules can withstand moderate bending during transportation and manipulation without any appreciable deterioration of their properties.  After gluing the fiber into the grooves, each strip is covered by a reflective aluminized Mylar tape to reduce photon losses and provide mechanical protection for the fibers as shown in Figure~\ref{Strip}-right.

At the read out end the fibers are diamond-cut and fixed in dedicated plastic end-pieces. These end-pieces also provide precise alignment of the fiber with the photo-sensor. The opposite end of each fiber is coated with aluminium by evaporation in vacuum forming a highly efficient reflecting mirror, which increases the light yield. Hamamatsu S12825-050P~\cite{Hamamatsu} Silicon PhotoMultipliers~(SiPMs) are used to collect the scintillation light from each fiber. Two SiPMs are used to read out one scintillation strip. Therefore, 32 photosensors are employed in one single module. Signals from the 32 SiPMs are sent to a Front-End electronics Board (FEB). Each FEB is mounted at the readout end of the modules and connected to the SiPMs via 1~mm thick coaxial cables.  Signal processing is performed in the FEB and readout for storage or higher-level triggering.

\subsection{Front-End Board Electronics} \label{subsec:FEB}

The CRT FEB is designed to serve 32 SiPMs from one module (16 scintillator strips). The FEB provides a bias voltage in the range of 40-90~V individually adjustable for each of the 32 SiPMs. At the same time it amplifies, shapes and digitizes the output signal of the photosensors. The FEB also provides the signal coincidence from each pair of SiPMs from the same scintillator strip with the possibility to trigger only on events that occurred in coincidence with another event in a different FEB or group of FEBs. Furthermore, the FEB is able to generate a timestamp taking an input reference with an accuracy of 1~ns. A detailed description of the FEB design and functionallity can be found in~\cite{Auger:2016vpo}.

\subsection{Scintillator Plane} \label{subsec:SciPla}

\begin{figure}[h!]
 \centering	
 \includegraphics[trim = 0mm 16mm 30mm 0mm, clip,width=0.75\linewidth]{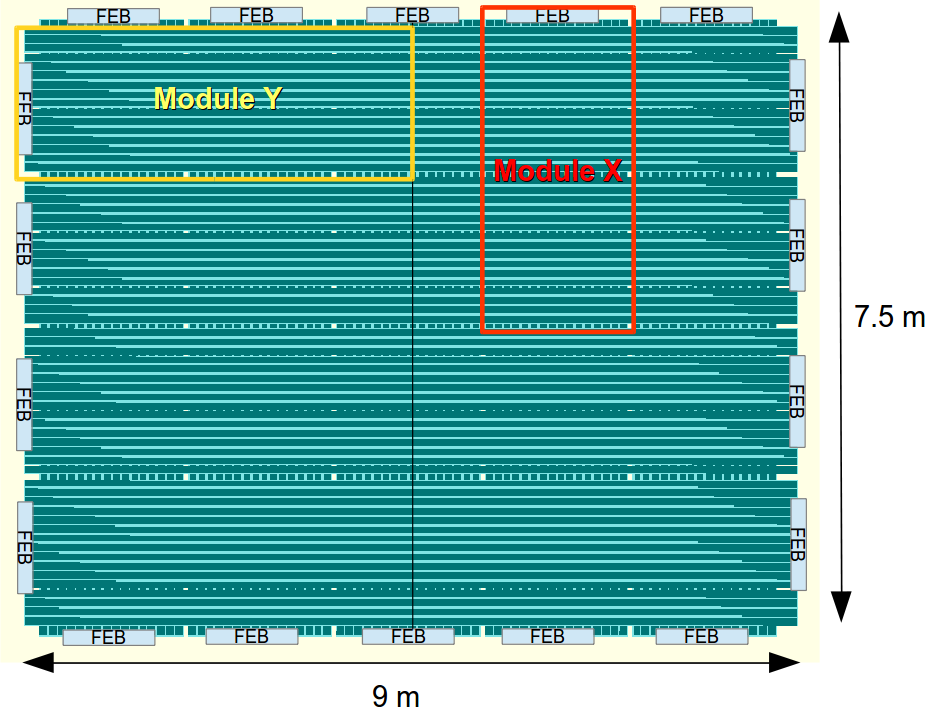}
 \caption{Cosmic Ray Tagger scintillator plane formed by two orthogonal layers of modules for X-Y coordinate reconstruction.}
 \label{fig:scinPlane}
 \end{figure}

A scintillator plane consists of several scintillator modules arranged in two different layers and orthogonal directions, as shown in Figure~\ref{fig:scinPlane}. In this configuration, when a charged particle crosses the plane, it fires at least one strip per module.

By using the relative intensities of the SiPMs corresponding to the same timestamp provided by the FEBs, the X-Y position of the particle crossing point can be computed (see Section~\ref{sec:Results}). Two different scintillator planes allow a crossing particle to be tagged, providing an "interaction vector". This information can then be extrapolated to the reconstruction algorithm of the LAr TPC, discriminating any related event within the volume associated to the interaction vector that could be mistakenly identified as part of a neutrino interaction event.

The total number of scintillator modules  to be built in Bern for the needs of the SBN program at Fermilab amounts to 215, for a total covering surface of 538~m$^2$.

\newpage
\section{Laboratory Tests} \label{sec:Results}

\begin{figure}[h!]
\centering	
\includegraphics[angle=90, width=0.85\linewidth]{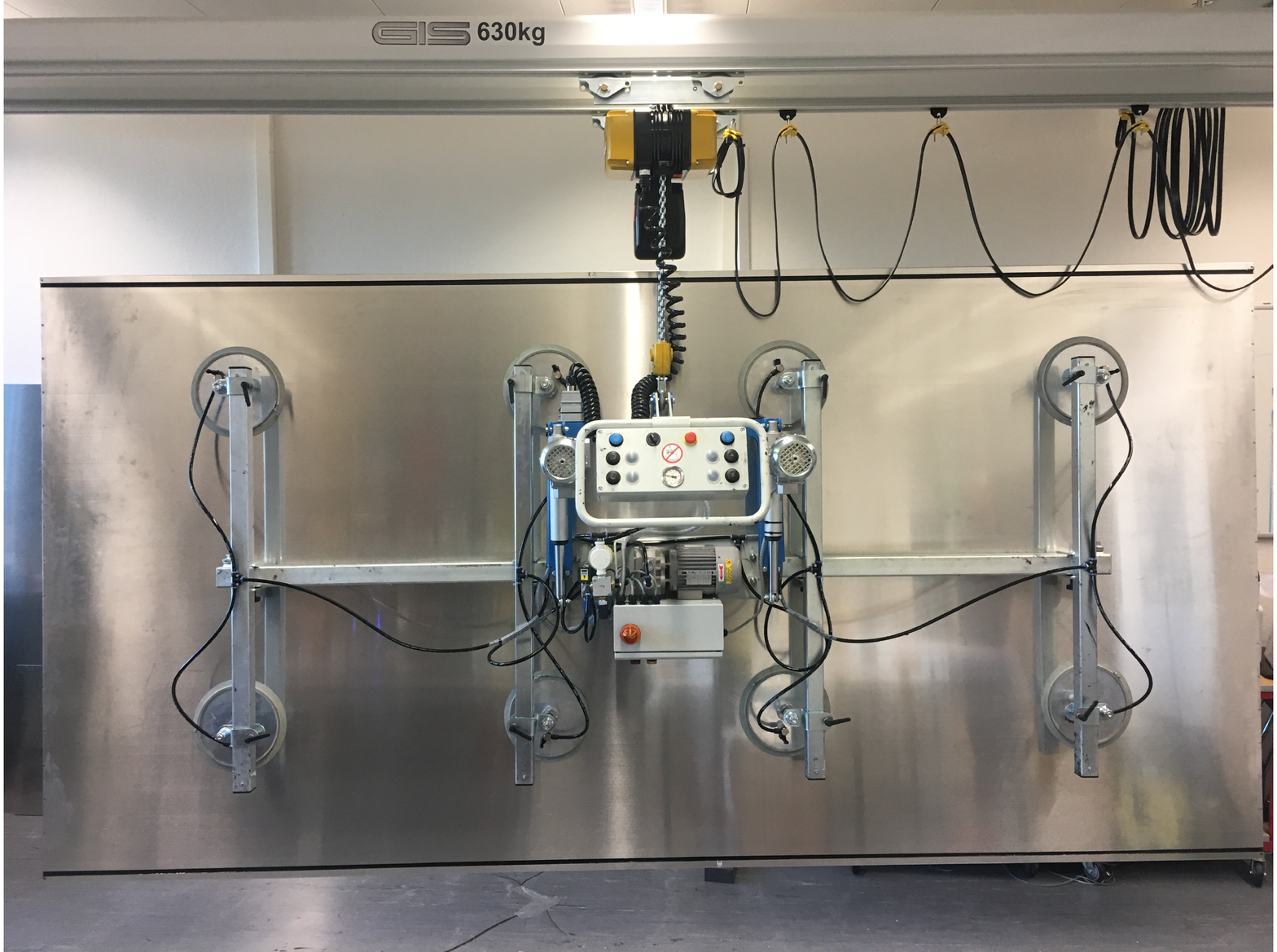}
\caption{Full-size 1.8 m x 4.1 m production module during testing at the University of Bern.}
\label{Module}
\end{figure}

The first full-size 1.7~m~$\times$~5~m prototype CRT module was assembled and tested at LHEP in July-August 2015. Different studies were performed with this test module providing useful information about the quality of the materials and the procedure for its construction. The same studies were conducted regularly on production modules (see Figure~\ref{Module}), obtaining similar performance results.

\begin{figure}[h!]
\centering
\includegraphics[width=0.98\linewidth]{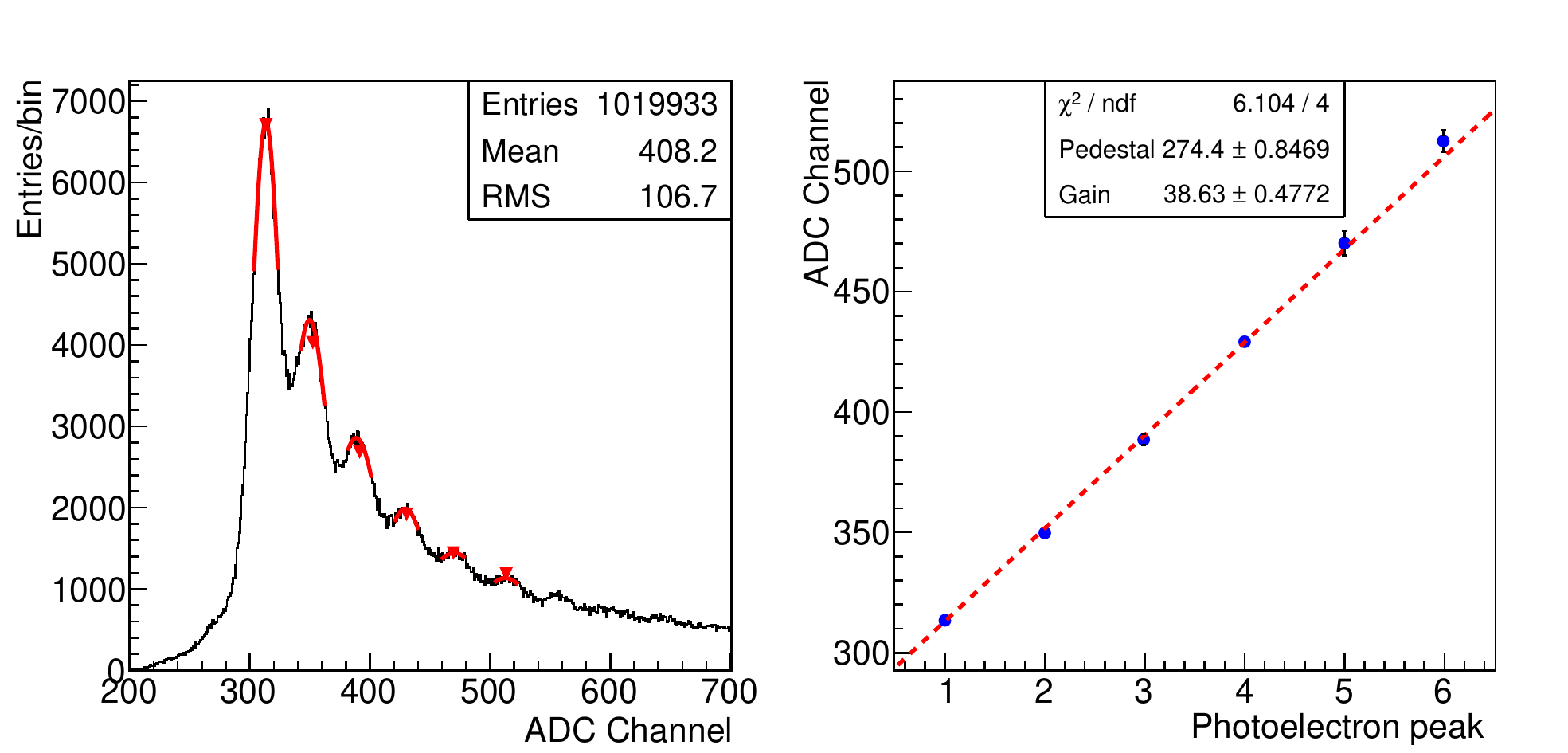}  
\caption{Left: SiPM spectrum obtained with ambient gamma rays. Individual photoelectron peaks are fitted to a Gaussian. Right: number of ADC counts produced by different number of photoelectrons, a linear fit allows to obtain the conversion gain factor.}
\label{fig:SiPMgain}
\end{figure}

Before data taking the conversion gain of each individual photosensor must be accurately determined. The SiPMs from each scintillator strip are calibrated using ambient background gamma events which cross the module. From these events the spectrum shown in Figure~\ref{fig:SiPMgain}-left is obtained, it corresponds to the digital signal associated to the avalanche charge produced by a different number of SiPM pixels. The conversion gain can be determined by fitting a Gaussian to the peaks and using the centroid position in a linear fit; an example is shown in Figure~\ref{fig:SiPMgain}-right.

\begin{figure}[h!]
\centering
\includegraphics[width=0.48\linewidth]{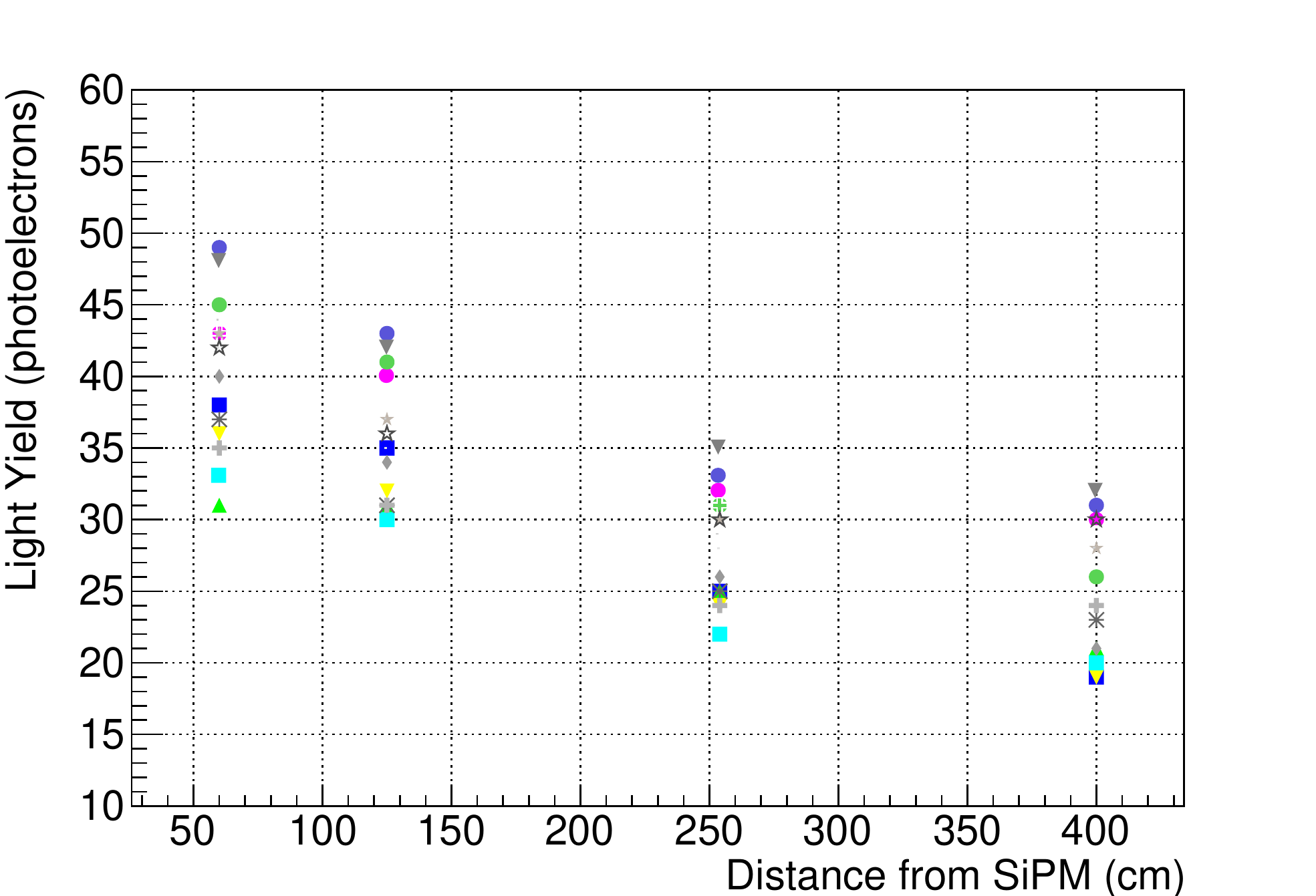}     \includegraphics[width=0.48\linewidth]{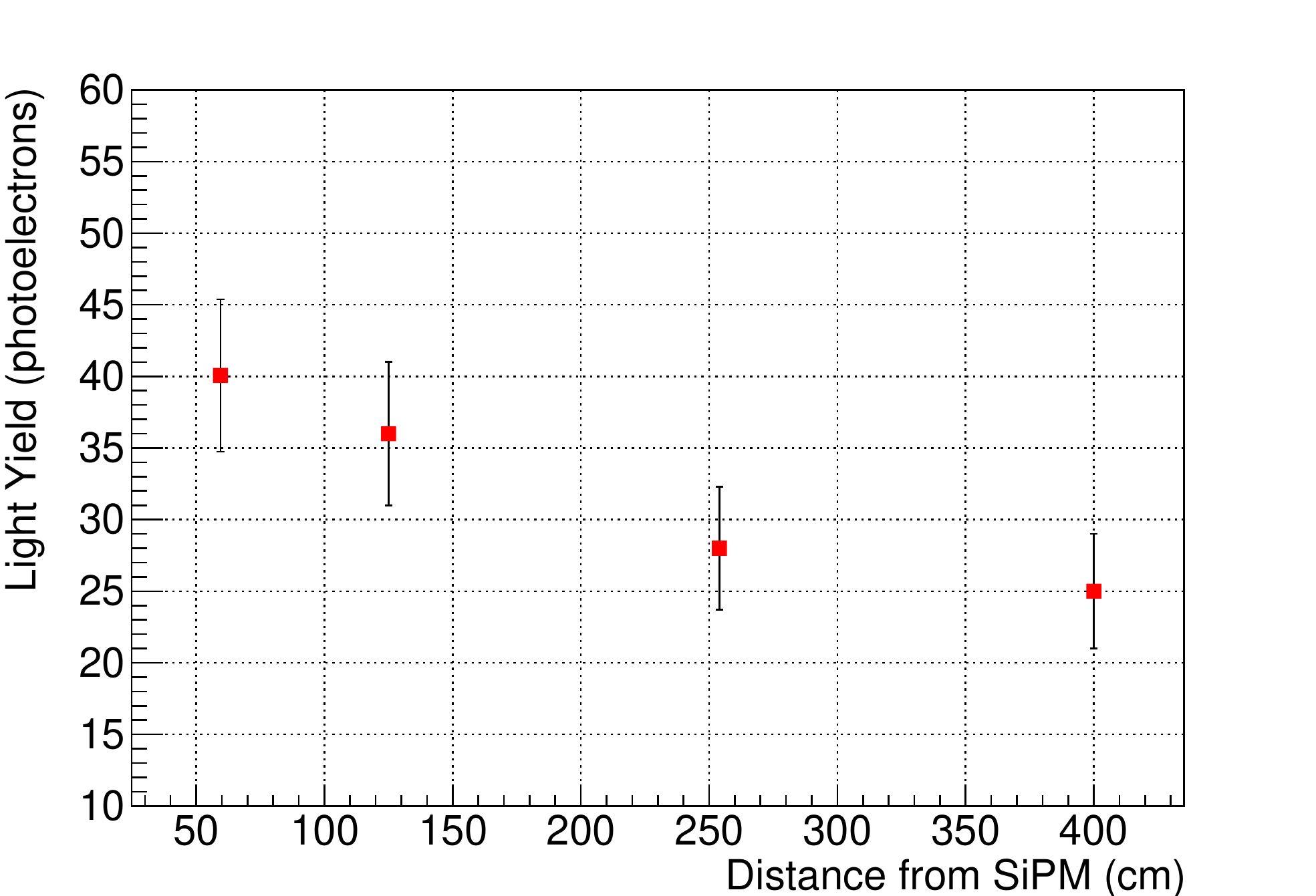}   
\caption{Scintillator strip response to a traversing relativistic muon as a function of distance from the readout end. Left: responses of individual strips. Right: mean and RMS of the data spread at each point.}
\label{fig-ly}
\end{figure}

The variation in light yield for a relativistic muon traversing the 16 scinctillator strips of the module at different distances from the readout end was also measured. The interaction distance was determined with the aid of an external muon telescope composed of two scintillator paddles with PMTs as photodetectors. The results for each of the 16 strips are shown in Figure~\ref{fig-ly}-left, with Figure~\ref{fig-ly}-right showing the mean and the RMS of the spread at each point.

As one can see, the light yield follows an exponential behaviour which can be fitted to extract the attenuation length of the scintillator. This was measured to be $6.88 \pm 0.01$~m, which is longer than the maximum module length of 4.1~m, hence ensuring sufficient light yield.

\begin{figure}[h!]
\centering
\includegraphics[width=0.7\linewidth]{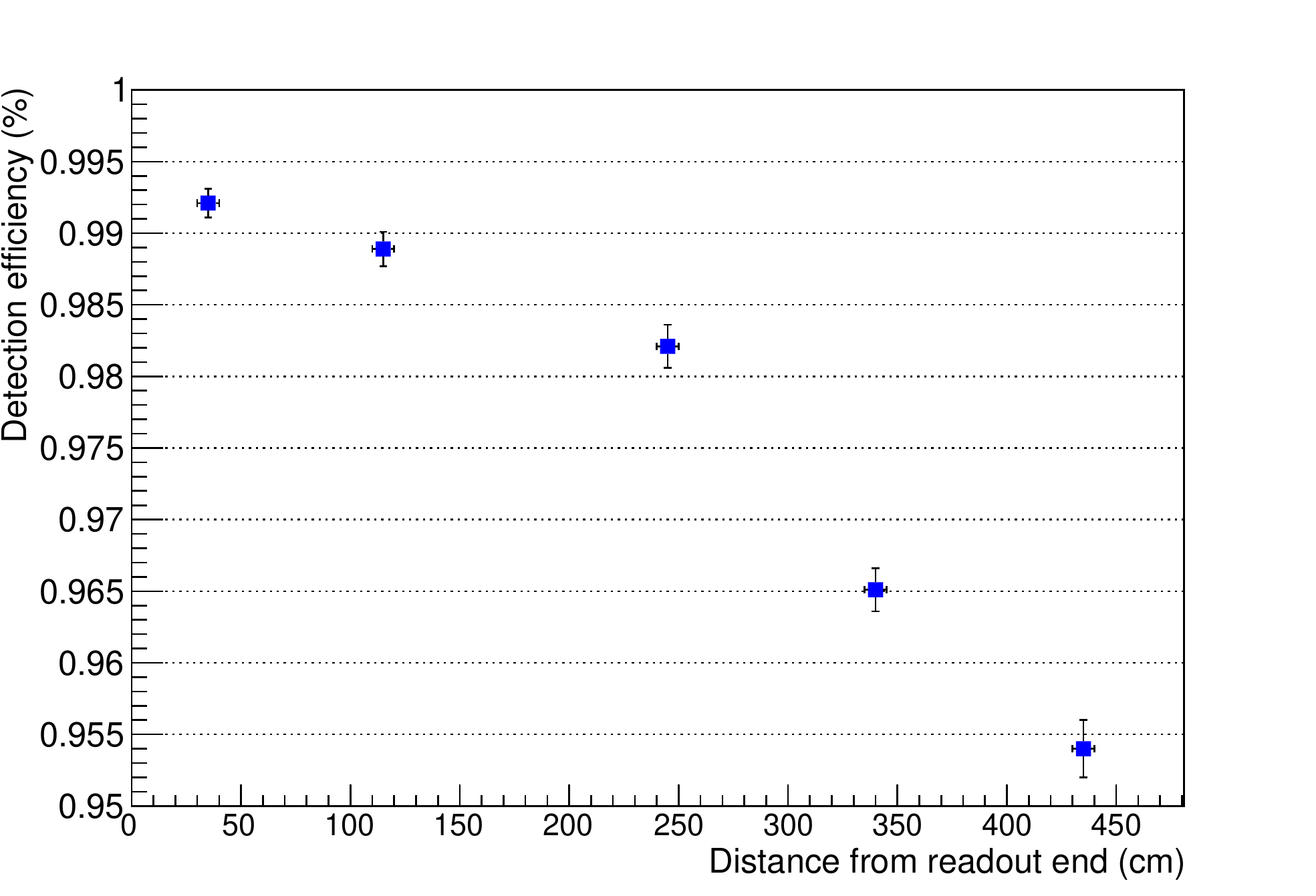}     
\caption{Detection efficiency for cosmic muons as a function of the interaction distance from the readout end.}
\label{fig-eff}
\end{figure}

The module detection efficiency for cosmic muons as a function of the distance from the readout end was measured using the same external muon telescope as in the previous tests. Detection efficiency is determined as the ratio between the number of muons detected by the telescope versus those detected by the module. As shown in Figure~\ref{fig-eff}, the efficiency is higher than 95\% across the plane.

\begin{figure}[h!]
\centering
\includegraphics[width=0.80\linewidth]{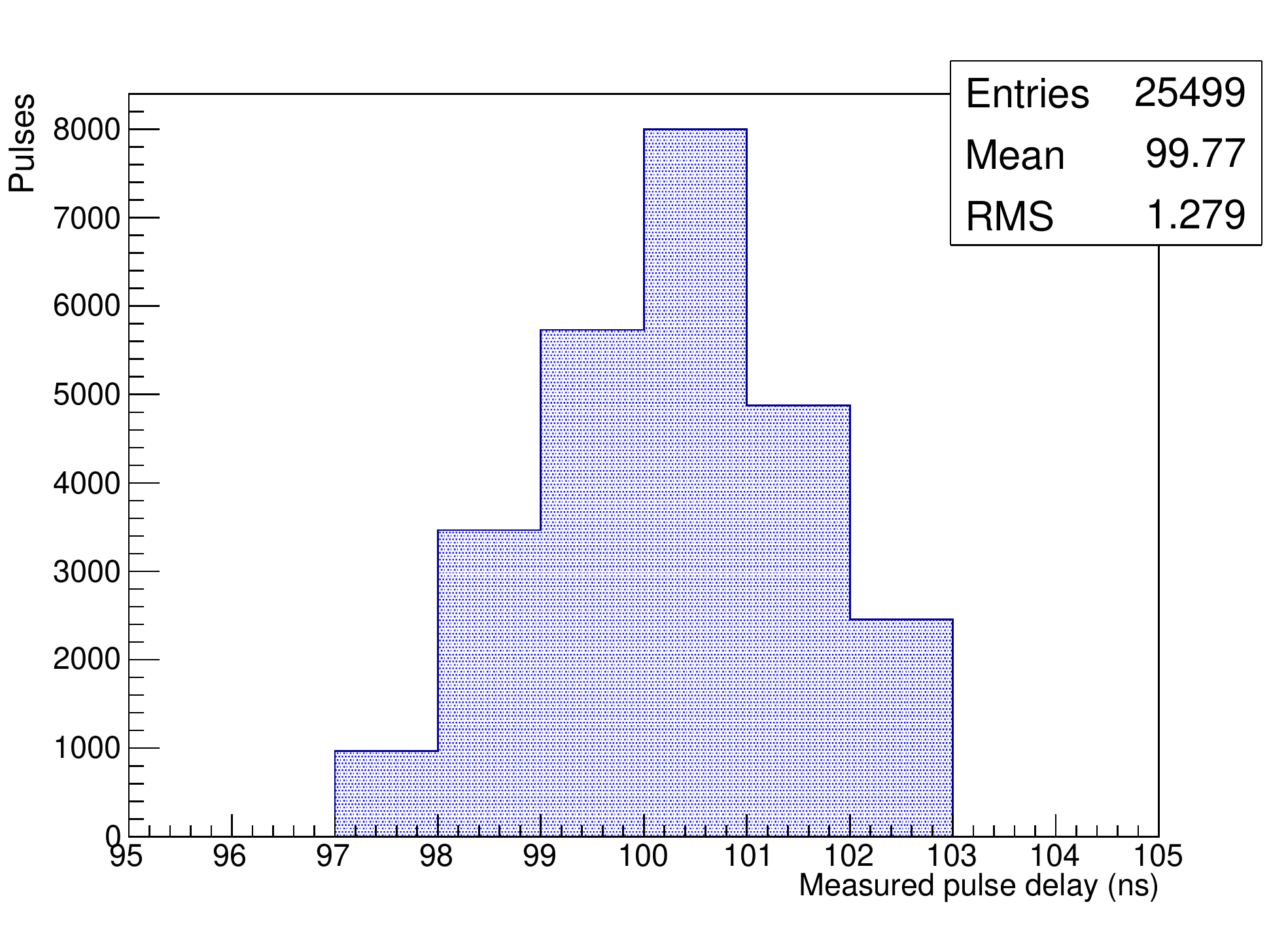}    
\caption{Accuracy of the FEB time stamp generator.}
\label{fig-tres1}
\end{figure}

The time resolution of the module is determined mainly by two factors: the resolution of the FEB time stamp generator and the spread of the photon's path inside the scintillator strip and the WLS fiber. The FEB time stamp generator is composed of a coarse counter working at the clock frequency of 250~MHz and a delay-chain interpolator~\cite{Auger:2016vpo}. For each event the FEB is capable of recording two independent time stamps w.r.t the positive flank on LEMO inputs. The resulting resolution is illustrated in Figure~\ref{fig-tres1} for events arrivng 100~ns after a reference signal; it is equal to 1.2~ns~RMS.

\begin{figure}[h!]
\centering
\includegraphics[trim = 15mm 10mm 20mm 25mm, clip,width=0.49\linewidth]{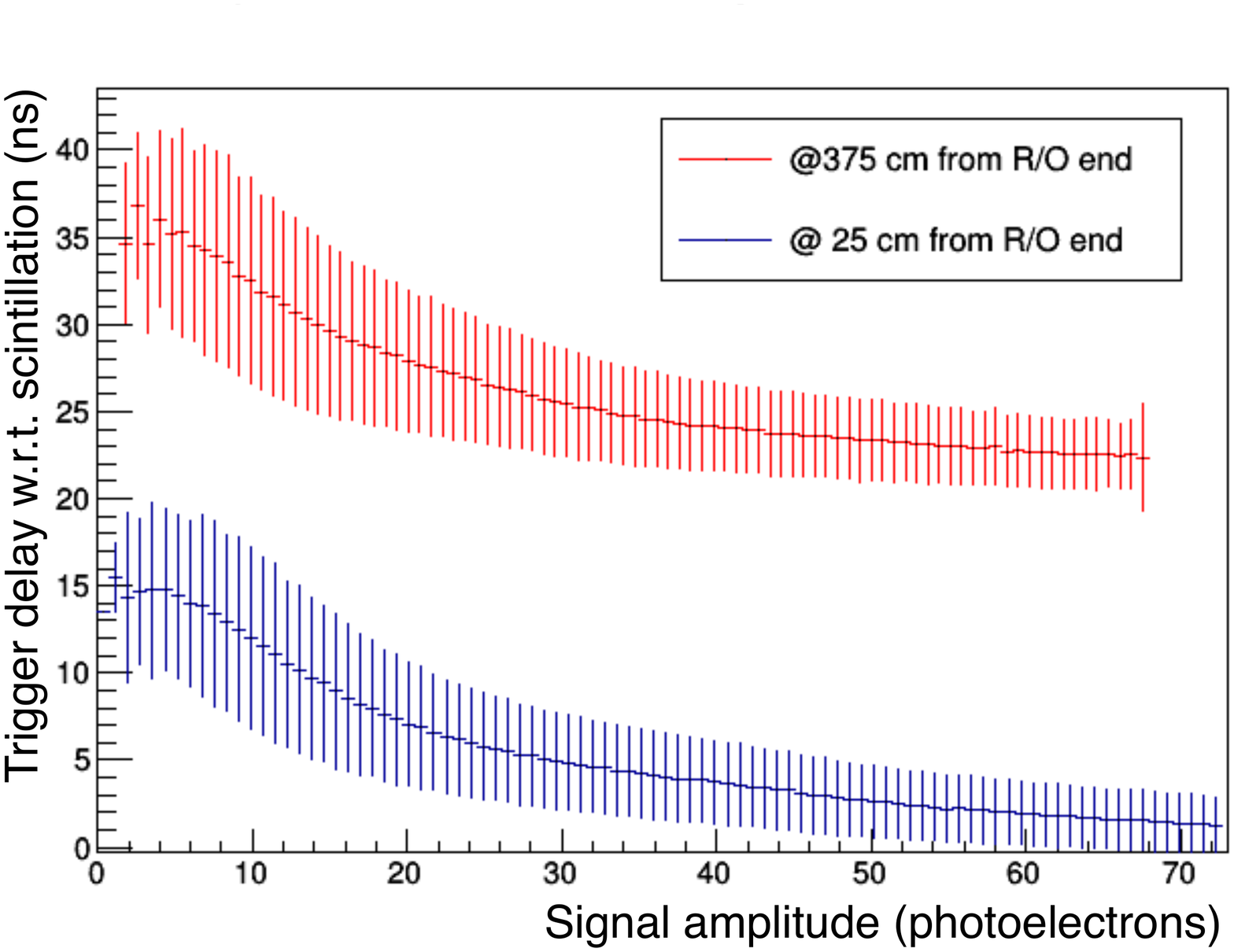}         
\includegraphics[trim = 15mm 10mm 20mm 25mm, clip,width=0.49\linewidth]{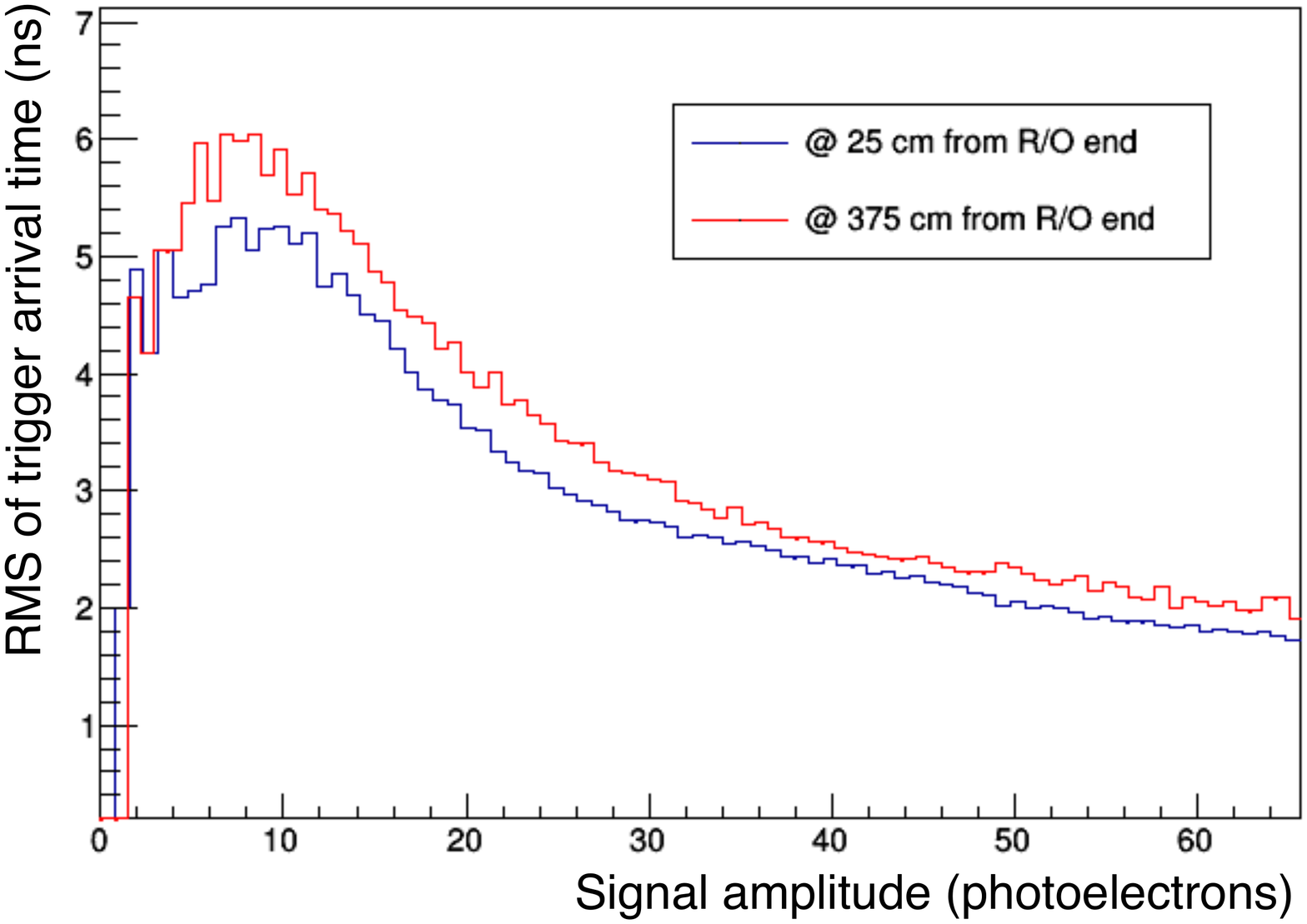}     
\caption{Arrival time (left) and its uncertainty (right) for the photons in CRT scintillator strips as a function of signal amplitude at a threshold of 2.5 photoelectrons. The effective photon propagation delay is 6.1$\pm$0.7~ns/m.}
\label{fig-tres3}
\end{figure}

The spread in the photon arrival time from the point where scintillation happened to the SiPM is measured with a 400~nm pulsed laser, set to 60~ps pulse width. The beam was illuminating the middle of the scintillator strip, between the two fibers. This measurement was made with the laser beam entering at 25~cm, and again at 375~cm from the SiPMs. The laser was triggered by a T0 reference signal, which was also sent to the FEB. The time registered at the FEB as event time represents the delay between the laser pulse and the signal at the SiPM. The beam was attenuated to a level where the SiPMs registered only a few photoelectrons. In Figure~\ref{fig-tres3} a summary of the results is shown. The two curves in Figure~\ref{fig-tres3}-left illustrate signal arrival delay for the two beam entry points. The resulting value for the effective photon propagation delay is 6.1$\pm$0.7 ns/m.  The dependence of the RMS on the signal amplitude for both 25 cm and 375 cm from the readout end is shown in Figure~\ref{fig-tres3}-right.

\begin{figure}[h!]
\centering
\includegraphics[width=0.49\linewidth]{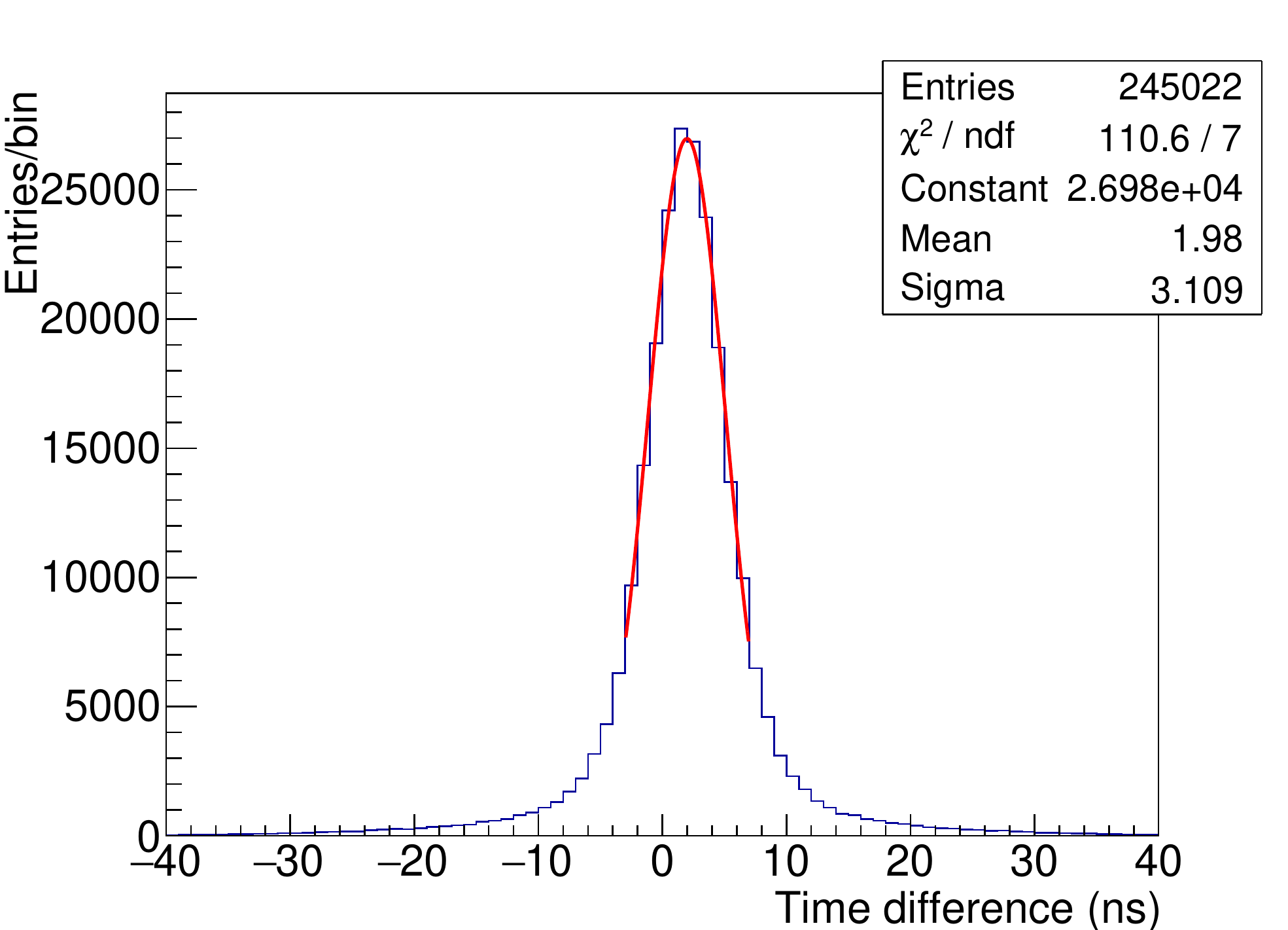}     
\includegraphics[width=0.49\linewidth]{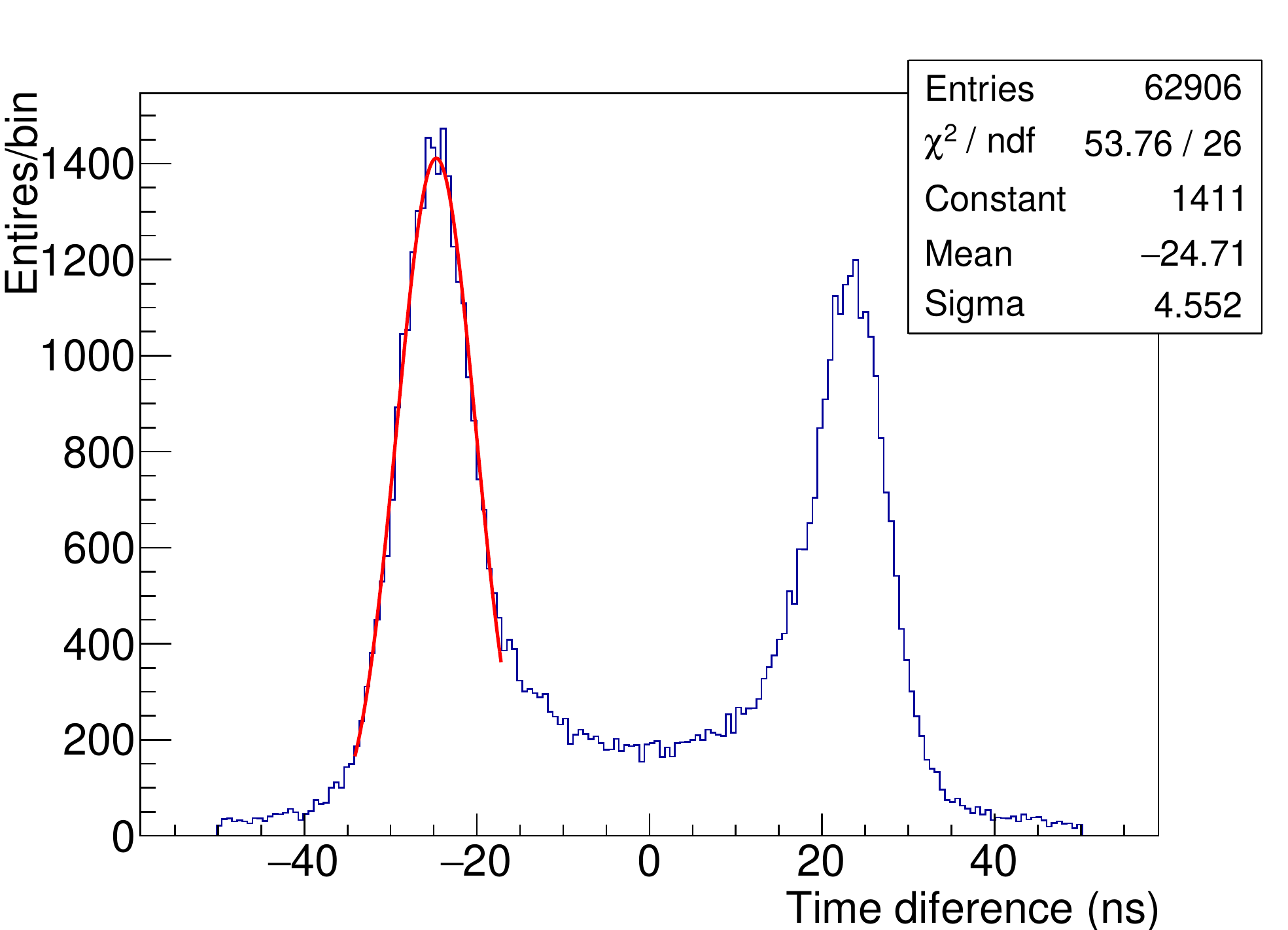}     
\caption{Trigger time difference distributions for two modules configuration with 0~cm (left) and 7.3~m (right) distance separation between them. Mean values correspond with delays introduced by the length of cabling used in each setup, 2~ns and 25~ns. The sigma values correspond to the trigger time resolution.}
\label{fig-timdiffmod}
\end{figure}

The difference in trigger times between two modules, in various configurations, was measured using crossing muons. Figure~\ref{fig-timdiffmod} shows results for the two modules triggering in coincidence; the left shows results from an adjoining module configuration, and the right with a separation of 7.3~m between modules. The mean values of the trigger time differences correspond with delays introduced by the length of cabling used in each setup, 2~ns and 25~ns respectively.

\begin{figure}[h!]
\centering
\includegraphics[width=0.49\linewidth]{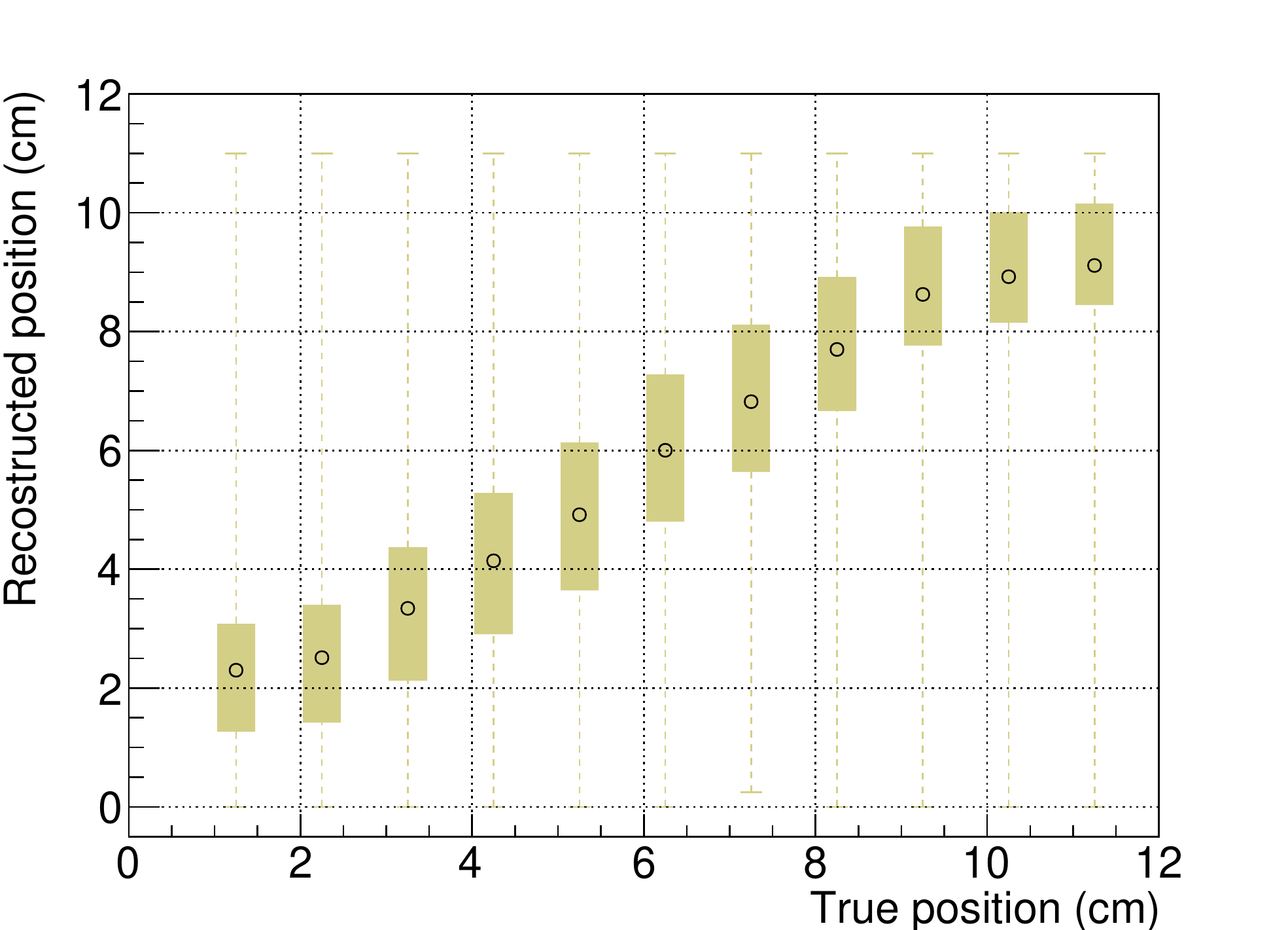}     
\includegraphics[width=0.49\linewidth]{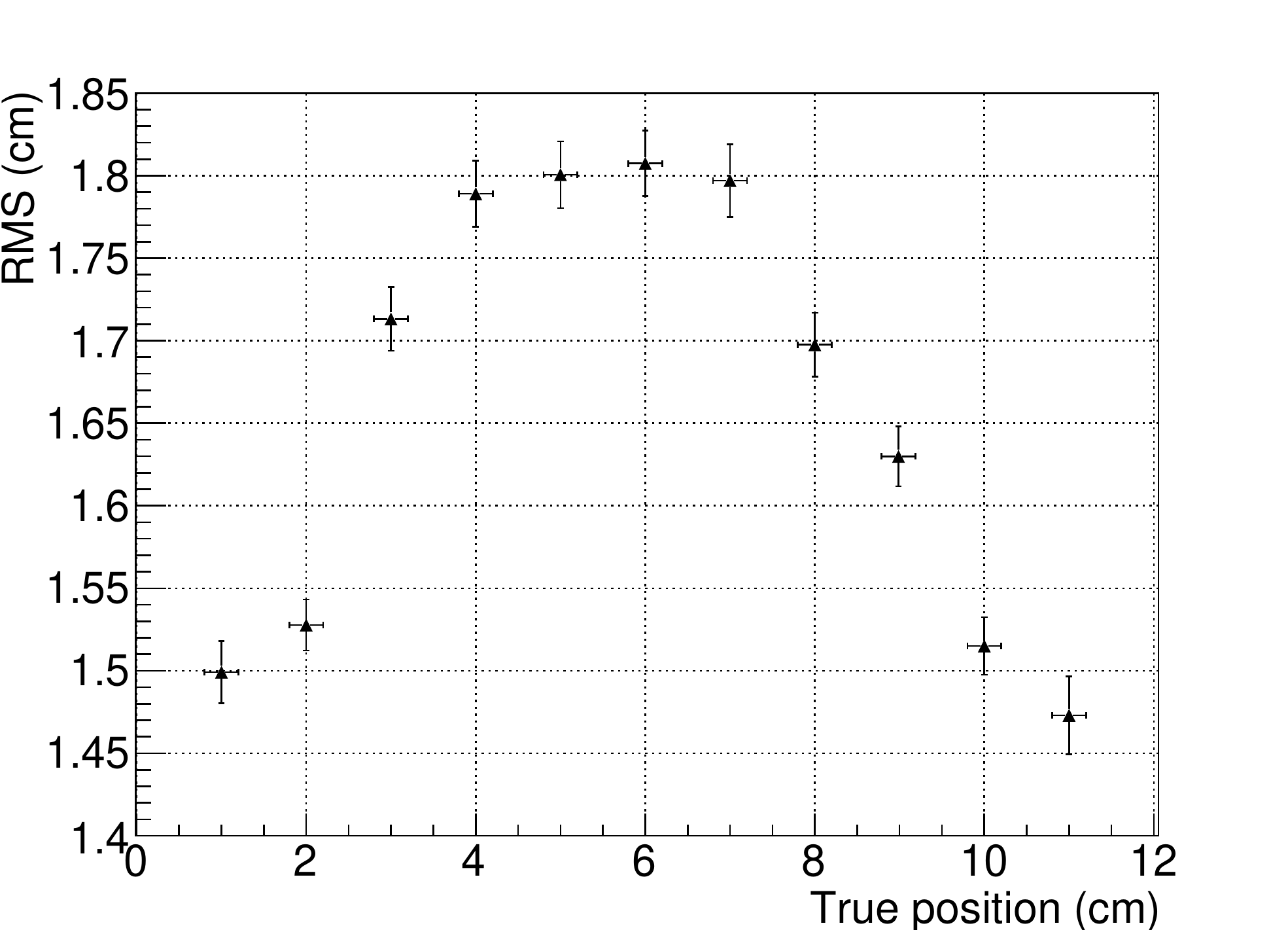}     
\caption{Left: position of hit derived as the weighted average of signals from left and right SiPMs. Right: RMS of the difference between the derived coordinate and the true one. The derived coordinate is corrected using the dependence in the left plot.}
\label{fig-cres}
\end{figure}

Each scintillator strip is equipped with two WLS fibers and two SiPMs, one on each edge.  This configuration enables the reconstruction of the position of a particle which hits the strip, by measuring the relative intensities registered by the SiPMs.  Measurements were made with cosmic muons entering normal to the plane at different transverse positions in the strip at 0.5~m from the readout end. The muon's true position was constrained with two scintillator counters working as a telescope in coincidence with the FEB of the module. Figure~\ref{fig-cres}-left shows the reconstructed hit coordinate versus its true position. Figure~\ref{fig-cres}-right shows the RMS of the difference between the reconstructed and true coordinates as a function of the true coordinate.

\newpage
\section{Summary} \label{sec:Summary}

We have presented a scintillator based Cosmic Ray Tagger (CRT) system, designed, built and characterised in Bern. The purpose of the CRT is to mitigate a dominant background of cosmic muons crossing the detectors of the Short Baseline Neutrino (SBN) program at Fermilab.

The SBN program will study $\nu_e$ appearance and $\nu_\mu$ disappearance, as well as neutrino argon cross sections. Tagging cosmic muons reduces the $\nu_e$-like backgrounds, improving the detectors' ability to identify and reconstruct neutrino interactions. 

The CRT system utilises a novel approach for construction of a scintillator-based tracking detector. Details of the construction, including materials, photosensors and other relevant technologies have been presented.
 
Characterisation of the individual modules of the CRT system has shown the spatial resolution to be 1.8~cm, with a timing resolution of 1~ns. The tagging efficiency was measured to be greater than 95\% across the the entire surface of each module. Quality assurance tests have shown the same performance across all modules built so far.

\vspace{6pt} 


\acknowledgments{We acknowledge financial support of the Swiss National Science Foundation. The Oxford group is  supported by the UK Science and Technology Facilities Council.}


\bibliographystyle{mdpi}

\renewcommand\bibname{References}


\end{document}